\documentclass{article}
\usepackage[english]{babel}
\usepackage{graphicx}
\usepackage{fixltx2e}
\usepackage{amsfonts}
\usepackage{amsmath}

\title{Automated Inference of Past Action Instances in Digital Investigations}
\author{Joshua I. James$^\dagger$, Pavel Gladyshev$^\ddagger$\\
Digital Forensic Investigation Research Laboratory\\
$^\dagger$Soonchunhyang University, Asan-si, Chungcheongnam-do, South Korea\\
$^\ddagger$University College Dublin, Belfield, Dublin 4, Ireland}
\date{}

\begin{document}
\maketitle


\begin{abstract}
\noindent As the amount of digital devices suspected of containing digital evidence increases, case backlogs for digital investigations are also increasing in many organizations. To ensure timely investigation of requests, this work proposes the use of signature-based methods for automated action instance approximation to automatically reconstruct past user activities within a compromised or suspect system. This work specifically explores how multiple instances of a user action may be detected using signature-based methods during a post-mortem digital forensic analysis. A system is formally defined as a set of objects, where a subset of objects may be altered on the occurrence of an action. A novel action-trace update time threshold is proposed that enables objects to be categorized by their respective update patterns over time. By integrating time into event reconstruction, the most recent action instance approximation as well as limited past instances of the action may be differentiated and their time values approximated. After the formal theory if signature-based event reconstruction is defined, a case study is given to evaluate the practicality of the proposed method.
\end{abstract}

\section{Introduction}
Since the definition of Digital Forensic Science at the Digital Forensics Research Workshop in 2001 \cite{Palmer2001}, the field has grown almost as dramatically as technology itself. As described by Casey \cite{Casey2009}, Digital Forensic Science is ``coming of age'', which not only brings about a maturation in the principal concepts of the field, but also an increased scrutiny against these principles and their lack of rigorous scientific backing. Digital forensic investigators are now finding themselves overwhelmed with the scale and quantity of cases, along with the pressure of increasingly restrictive standards \cite{Garfinkel2010}. This combination translates into a consistently increasing number of delayed, or even neglected, cases.

Gogolin \cite{Gogolin2010} states that in Michigan, USA, ``50\% or more of [Law Enforcement's] cases have a digital component, and most agencies report that this number is increasing''. This is not surprising given the rapid adoption and evolution of technology worldwide on a business as well as personal level within the last 10 years alone. As technology becomes more a part of everyone's life, it is natural that more evidence in investigations will be found in digital form. The issue is that many law enforcement agencies are not currently well positioned to handle an ever-increasing amount of data using traditional digital forensic techniques. Casey, et al. \cite{Casey2009a} observes ``few [digital forensic laboratories] can still afford to create a forensic duplicate of every piece of media and perform an in-depth forensic examination of all data on those media''. Even though the field of digital forensics has been advancing rapidly, the backlog for digital investigations has continued to increase. Currently in the United States there are reports of backlogs from 12 to 18 months \cite{Raasch2010}, and in some cases ``approaching or exceeding 2 years'' \cite{Gogolin2010}. In 2004 the United Kingdom digital crime investigation backlog was 6 to 12 months \cite{BBC2004}, and rose to 18 to 24 months in 2009 \cite{InfoSecurity2009} before being improved through a number of policy, case prioritization and evidence outsourcing initiatives \cite{Kohtz2011}. Since technological advancement for personal and business use shows little sign of slowing, data and backlog growth will continue unless law enforcement, and the legal system in general, move towards the use and acceptance of verifiable, highly-automated solutions during the digital investigation process.

In an effort to provide faster data to knowledge acquisition for the digital investigator, this research proposes the use of signature-based methods for the automated analysis of actions that happened in a given digital system. It shows that the state of low-level artifacts in a suspect system may be automatically observed and correlated to higher-level actions using signature-based methods that take into account measured trace update thresholds, rather than the assumption of an immediate trace updates on the occurrence of an action. The result is a fast and detailed reconstruction of action instances that could be applied during the triage, preliminary, and in-depth analysis phases of an investigation.

\section{Related Work}

\subsection{State Machine Analysis}
Formal analysis is a method to formally represent a system and analyze certain scenarios based on this formal model. Several works have proposed modeling a computer system formally as a finite state machine (FSM) \cite{Gladyshev2004,Carrier2006,Arasteh2007}, which allows event reconstruction to be reduced to a state-space exploration problem. FSM models have been applied to model suspect systems in various ways. For example, Carrier \cite{Carrier2006} proposed a computer history model that groups the system into primitive (lowest level) and complex (causing multiple primitive or complex events) events. An investigator then formulates hypotheses of events that are tested against the created computer history model in order to support or refute the hypothesis. Gladyshev and Patel \cite{Gladyshev2004} took a different approach by modeling the suspect system as an FSM where only the final state is known. The state-space is then back traced in order to find all possible scenarios that could have resulted in the final, observed state.

The issue, and benefit, of creating a state machine model of a suspect system is that all possible states and transitions of the system are considered. However, in real-world systems the state-space of even the simplest system becomes computationally impractical. Even with efforts to reduce the possible state-space, as described by James, et al. \cite{James2010a}, computational modeling of a real-world system for analysis currently requires too much abstraction to be practical.

\subsection{Computer Profiling}
The work on computer profiling presented in Marrington, et al. \cite{Marrington2010} attempts to generate a computer usage profile that ``{\dots} allows a human examiner to make an informed decision regarding the likely value of the computer system to an investigation before undertaking a detailed manual forensic examination''. In this work an abstracted object model is used to classify objects observed in a suspect system. Observed objects are categorized as particular object types, such as system, principle (people/groups), application or content data. Relationships between these objects are then determined. These relationships provide insight into the logic of the system, and allow for the identification of indirect relationships between objects that were otherwise thought to be unrelated. By examining an object of interest the relation of other objects may be found. Times and events -- defined as recorded, inferred and unknown types -- are found, and are associated with their corresponding objects, where possible. The overall computer profile is then represented by these object, relationship and event connections. Hypotheses about a computer system and its history can then be formulated and tested based on the derived computer profile.

By concentrating on an informational rather than a computational finite state machine model, this method does increase practicality compared to methods previously described. In general, states are defined at a more abstract (object) level, and are based on observed evidence. Because of this, creation of the informational model is less computationally intensive as not all possible combinations of past states must be considered. The drawback is this model represents suspect objects and their relations, but makes no conclusion about what exactly these relations mean. The investigator is still left to manual hypothesis generation and testing, where some previously mentioned methods attempt to automatically present possible hypothesis as well as test the hypotheses within the created models.

\subsection{File System Activity Analysis}
One probabilistic method of file system activity analysis has been presented by Khan and Wakeman \cite{Khan2006}. In their work, neural networks are used to learn and detect application ``footprints''. Traces that were created on a disk (usually file creation and manipulation) by a particular application of interest were fed into the neural network in the order in which they were accessed to learn the update time-span relationships and file-system manipulation patterns of the application.

This method is highly probabilistic, where the neural network is able to calculate the likelihood of an observation matching a previously derived model (footprint). Khan, et al. \cite{Khan2007} showed that neural networks trained on specific application trace creation variables show relatively good results in determining and differentiating one application's footprint from another. One issue -- that was also discussed by the authors -- is that these systems need a very large amount of training data to be reliable. The training data needed, manual variable selection and separate neural networks per application make this method less practically feasible. Also, when comparing the application footprint of this method to the use of signatures that encode multiple object update behaviors described in James, et al. \cite{James2010b}, the differences in models suggest that learned signatures lack some specificity that could provide more event information for use in reconstruction.

\section{Detection of Action Instances}
Unlike previous works, this work focuses on the automated extraction of user action instance hypotheses. An action is any event external to the system that is the direct cause of a process. An action is the farthest point at which a happened event can be inferred. Opening a program by clicking an icon, for example, is an external activity conducted by a user that manipulates the state of the system. This work submits the hypothesis that multiple past executions of an action may be inferred by analyzing sporadically updated artifacts associated with the given action instance. In this work the detection of past action instances using signature-based methods that includes artifact update time-spans will be described, and practically demonstrated in a simple case study.

\subsection{Theory of Action Instance Reconstruction}
When a user interacts with a system, actions cause changes in the state of the system. Given the deterministic nature of computer systems, a certain action will consistently cause the same changes within the system each time the action takes place if the system is in the exact same starting state. A user clicking on a program's icon, for example, will consistently execute the program. The program must access specific files to load, which in turn updates files, file metadata, log entries, etc. This work will focus on file meta-data, and specifically file access, modified and created time stamps.

\begin{figure}
	\caption{Causal chain of an action causing a process that causes trace creation.}
	\centering
	\includegraphics[width=0.9\linewidth]{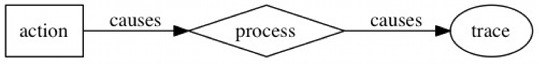}
	\label{fig:actionTraceChain}
\end{figure}

A causal relation between actions and object updates (trace creation) via process execution can be determined because of causal chaining, where an action causes the process and the process causes the update of corresponding objects (Fig. \ref{fig:actionTraceChain}).

\begin{figure}
	\caption{Back-tracing the causal chain from the observed trace to determine the corresponding action that caused the trace.}
	\centering
	\includegraphics[width=0.9\linewidth]{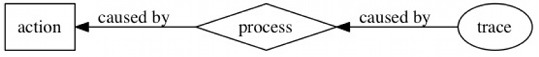}
	\label{fig:traceActionChain}
\end{figure}

Causal links can be traced back from the final observed state to determine all possible beginning actions that could result in such a final state \cite{Menzies2008}. In this case, the observation of a resulting trace may be related back to the process that caused it, which may in turn be related back to the action that caused the process (Fig. \ref{fig:traceActionChain}).

A system contains a finite set of objects, $O$, where each object in $O$ may be defined in terms of associated access ($ta$), modified ($tm$) and created ($tc$) time stamps:

\begin{description}
\item $o = (ta, tm, tc)$
\end{description}

The objects and time stamps in a system can be described as follows:

\begin{description}
\item $t = (\tau)$, where t is a time stamp and $\tau$ is the time value of the time stamp
\item $O = \{o_1, o_2, o_3{\dots}\}$
\item $Tm = \{tm_1, tm_2, tm_3 {\dots}\}$
\item $Ta = \{ta_1, ta_2, ta_3 {\dots}\}$
\item $Tc = \{tc_1, tc_2, tc_3 {\dots}\}$
\item $T = Tm \cup Ta \cup Tc$
\end{description}

\begin{description}
\item An action is defined as:
\item $a = (Ma,Da,Oa,af)$
\item where:
\end{description}

\begin{itemize}
\renewcommand{\labelitemi}{$\bullet$}
\renewcommand{\labelitemii}{$\diamond$}
\item $Ma$ is an action that modifies timestamps to the current time plus some random period of time $(\tau + \varDelta\tau)$
\item $Da$ is an action that sets timestamps to a default value
\begin{itemize}
\item $Da = \{ (t,\tau) \}$, where $t$ is the timestamp, $\tau$ is the default value
\end{itemize}
\item $Oa$ is a collection of objects the action creates if they are not present
\item $af$ is a function which takes in a set of objects and returns another set of objects produced from the original
\begin{itemize}
\item $O' = af(O)$
\end{itemize}
\end{itemize}

As defined, an action may update time stamps. Since the time stamp update period is not instantaneous, the update will happen at some random interval after the action. This update takes place at a random delta after the time of the action. The action will also create an object if it does not exist. If objects, and their corresponding timestamps have been destroyed, these timestamps are not updated. Created objects may have timestamps set to default values that may possibly be sometime before the time the action took place. For example, software installation and backup recovery actions could produce objects with time stamps that are before the installation and backup actions occurred. Finally, an action function exists that accepts a set of objects as an input, and returns a modified set of objects produced from the original.

\begin{description}
\item The set of all actions is defined as:
\item $A = \{a_1,a_2,a_3 \dots\}$
\end{description}
Actions happen at a particular point in time. An instance of an action is defined as:
\begin{description}
\item $I = \{(a,\tau)|a \in A,\tau \in \mathbb{R}\}$
\item where:
\end{description}

\begin{itemize}
\renewcommand{\labelitemi}{$\bullet$}
\renewcommand{\labelitemii}{$\diamond$}
\item $\tau$ is the time of the action instance occurring
\end{itemize}

In the proposed model, the set of actions is defined such that each timestamp is a result of some action. This simplified system model is defined as:
\begin{description}
\item $\forall t \in T, \exists i \in I, (t.\tau = i.\tau + \varDelta \tau)\vee\\(\exists d \in i.a.Da, (d.t = t) \wedge(t.\tau = d.\tau))$
\end{description}

This model states that for all time stamps there either exists an action instance where the current time of the time stamp equals the time of the action instance plus some random period of time, or the time stamp is equal to a default time stamp.

In the model so far defined, a single action deterministically causes modification or creation to a set of objects. However, in a real-world system a single execution of the action can have multiple paths through a program. An action with multiple paths through a program is defined as:

\begin{description}
\item $aa = \{a_1,a_2,a_3,a_4\}$
\end{description}

Each path through a program will cause different modifications. The result is that not all time stamps may be updated with every execution, depending on the path.

The event reconstruction approach adopted in this work seeks to recover a sequence of action instances ($i$) where:

\begin{description}
\item $i = (a, \tau)$ 
\item $I = \{i_1,i_2,i_3,i_4 \dots\}$
\end{description}
And where multiple action instances may have an effect on a single object.
\begin{description}
\item $O' = in.aa.af( in-1.aa.af \\(\dots i2.aa.af( i1.aa.af (O))\dots ))$
\end{description}
A reconstruction function ($ISR$) is defined as:
\begin{description} 
\item $I' = ISR(O')$
\item where:
\end{description}
\begin{itemize}
\renewcommand{\labelitemi}{$\bullet$}
\renewcommand{\labelitemii}{$\diamond$}
\item $I' \subseteq I$
\end{itemize}

In this case $I' = I$ is impossible. Correctness is such that $i$ in $I'$ implies $i$ in $I$. However, the opposite is not necessarily always true.

Since the object update delay is random, it cannot be said when the action occurred. Likewise, different action instances $(a_1, \tau_1)$ and $(a_2, \tau_2)$ may be executed at different times but because delta is different for each action, the final time stamp may be the same. For example, if:

\begin{description}
\item $i_1 = (a_1, \tau_1)$
\item $i_2 = (a_2, \tau_2)$
\item $\tau_1 \ne \tau_2$
\end{description}

It is possible that $O' = i_1.a.af(O)$ and $O' = i_2.a.af(O)$ if deltas for each action instance are different. Therefore, a function that uniquely identifies multiple action instances that acted upon the final observed object is impossible; however, it is possible to recover a subset of possible action instances.

Provided delta is random, the exact time of an action instance cannot be found, but the time of an action instance may be statistically approximated. With a sample of the time intervals in which an action instance takes to update objects $(\varDelta\tau)$ a distribution for the action instance update threshold can be created. Given a probability distribution of delta, the time of the action instance can be time-bound, and probabilistically approximated. The threshold for an action instance execution is defined as:

\begin{description} 
\item $\theta = \Phi (\varDelta\tau_1,\varDelta\tau_2,\varDelta\tau_3,\varDelta\tau_4 {\dots})$
\item where:
\end{description}
\begin{itemize}
\renewcommand{\labelitemi}{$\bullet$}
\renewcommand{\labelitemii}{$\diamond$}
\item $\theta $ is the maximum action instance execution threshold
\item $\Phi $ is a probability density function that accepts the set of $\varDelta\tau$ for a single action instance
\end{itemize}
A function that identifies the exact time of the action instance cannot be defined, but limits on the duration of the action instance are restricted to within $\theta$ from the set of time stamp values.

\begin{description}
\item $0 \leq \varDelta\tau \leq \theta$
\item
\item The definition of an action instance must be updated to account for the execution threshold:
\item $i = (a, \tau , \theta)$ 
\end{description}

The aim the proposed algorithm is to recover a subset of action instance approximations where each approximation consists of an action that happened and an approximation of the time in which this action happened. An action instance approximation ($ia$) is defined as:

\begin{description}
\item $ia = (a, \lambda\tau)$
\item where:
\end{description}
\begin{itemize}
\renewcommand{\labelitemi}{$\bullet$}
\renewcommand{\labelitemii}{$\diamond$}
\item $\lambda\tau$ is the time interval in which the instance must have occurred in the form of a double containing two time values $[\tau_1, \tau_2]$
\begin{itemize}
\item $\tau_1$ is the start time of the interval and $\tau_2$ is the end time of the interval
\item $\tau_1$ and/or $\tau_2$ may be \textit{null} denoting no limit
\end{itemize}
\end{itemize}

\begin{description} 
\item $Ia = \{ia_1,ia_2,ia_3 {\dots}\}$
\end{description}

To recover a subset of action instance approximations ($Ia$), first the function $AIA$ is defined that returns a time interval ($\lambda\tau $) from a given time value ($\tau$) and an action instance execution threshold ($\theta $).

\begin{description} 
\item $\lambda\tau = AIA(\tau, \theta )$
\end{description}

The function $AIA$ returns double $[\tau_1, \tau_2]$, where $\tau_1$ is calculated by subtracting $\theta $ from $\tau$, and $\tau_2 = \tau$.

The function $ISR$ is defined that returns the set of possible action instance approximations ($Ia$) from the final observed state $O'$, and the set of all action instances $I$.

\begin{description}
\item $Ia = ISR(O', I)$
\end{description}

In $ISR$, for each time stamp $(O'.o.t)$ in the input set $O'$, and for each update instance $(i.a.Ma)$ in the input set $I$ that the time stamp is a member of, get the action path $(I.i.aa.a)$ where the update instance set contained only the single time stamp $(I.i.aa.a.Ma = \{O'.o.t\})$, and the result of $AIA(O'.o.t., I.i.\theta )$. The result is a set of all possibly executed actions that is a subset of $I$, and a set of instance approximation intervals per action instance.

\subsection{Signatures of Action Instances}
This work submits the hypothesis that signature based methods may be used to automate the trace observation, action inference and action instance approximation tasks. For the task of observation, object time stamps and their relation to the action instance must be known. For action instance execution approximation, the action instance execution threshold is required, where an unknown (\textit{null}) value equals any time in the past. And for the inference task, understanding of the underlying relation between the observed facts and the inferred conclusion is required. Knowledge of the system may be encoded as a trace update consistency-checking function. From this, a signature is defined as:

\begin{description} 
\item $S = \{Ti,\theta,cm\}$
\item where:
\end{description}
\begin{itemize}
\renewcommand{\labelitemi}{$\bullet$}
\renewcommand{\labelitemii}{$\diamond$}
\item $Ti$ is the set of all object time stamps associated with the action instance in the form of an object-trace double $[o, t]$
\begin{itemize}
\item $Ti = \{t|t \in i.a.Ma \}$
\end{itemize}
\item $\theta$ is the maximum action instance execution threshold
\item $cm$ is the update consistency checking function particular to the category of object update patterns
\end{itemize}

A method for the derivation of time stamps related to a particular action instance has been described by James \cite{James2011}. This method allows for the determination of the set $Ti$ related to a particular action instance. The derivation of the object update threshold ($\theta$) for a particular action instance, which will be described. This section, however, will focus on the update consistency checking function ($cm$), and the definition of three main time stamp related update patterns.

\subsubsection{Core Object Update Consistency}
Core object time stamps are defined as \textit{a subset of time stamps $S\textsubscript{core}$ in $T$ that are updated to the current value of the system clock on the occurrence of each execution of a single, specific action}.

All of the time stamps in a Core set are said to be in the `always updated' time stamp category. Using this definition, if any trace in a Core set is observed then it can be inferred that the action instance must have happened since the artifact relates to one, and only one, action.

Also, since Core time stamps are `always updated', it is expected that each time stamp will be within a certain time range of each other depending on the particular object update threshold of the action.

An example of a Core trace would be a configuration file that is always modified when its related program, Program X, is closed. If the configuration file were only modified when Program X is closed, the modification time stamp of the configuration file would be a Core trace for the action ``Close Program X''.

From this definition, an object time stamp update consistency function ($CoreTest$) can be derived to test whether each object update conforms to the Core signature category. In the case of Core, if each trace has been updated within $\theta $, then the execution time for the action instance can be time-bound before the oldest time in the array.

First, a function $getTraceStates$ is defined to return the state of time stamps of all objects defined in S. For each object specified in the signature, add the object, time stamp and time stamp value to the array $TraceStates$.

\begin{description} 
\item \textbf{function} $getTraceStates(O',S)$
\item array $TraceStates$
\item foreach $S.Ti.o \in O'$
\item $let TraceStates = TraceStates + [S.Ti[o,t], O'.o.t.\tau]$
\item return $TraceStates$
\end{description}

Next, the function $CoreTest$ may be defined that accepts an object update threshold and the $TraceStates$ array. First the $TraceStates$ array is sorted based on the time stamp values, where element $0$ is the oldest and $n-1$ is the newest (most recent) time stamp value. If the oldest time stamp value in $TraceStates$ plus the object update threshold is less than the most recent time stamp value in $TraceStates$, then the Core traces are not consistent. If the oldest time stamp value plus the object update threshold is greater than the most recent object update value, then the Core traces are considered consistent. The function $CoreTraces$ return the array $detected$, which is a single element array containing a double with the oldest and most recent time stamps in $TraceStates$.

\begin{description} 
\item \textbf{function} $CoreTest(\theta, TraceStates)$
\item sort $TraceStates$
\item if $(TraceStates[0,0] + \theta < TraceStates[n-1])$
\item return $null$
\item else
\item return $detected[TraceStates[0,0], TraceStates[n-1]]$
\end{description}

\subsubsection{Supporting Object Update Consistency}
Supporting object time stamps are defined as \textit{a subset of time stamps $S\textsubscript{support}$ in $T$ that may or may not be updated to the current value of the system clock on the occurrence of each execution of a single, particular action instance, but that will only be updated by a single, particular action}.

Supporting object time stamps are in the `irregularly updated' time stamp category. However, similar to Core signatures, if any trace in a supporting signature is detected, then it can be inferred that the action instance must have happened since the trace also relates to one, and only one, action.

A time stamp can be irregularly updated if, for example, a file is cached in memory after the first execution of an action. If the file data cached in memory, rather than the file on disk, is accessed on the next execution of the action instance then the trace update will not be observable on the disk. In this case the original file's meta-data on disk would not be updated on the execution of the second action instance.

From this definition, an object time stamp update consistency function ($SupportTest$) can be derived to test whether each trace conforms to the supporting signature category. In the case of supporting, if each trace has been updated within $\theta $, then the execution time for the action instance can be approximated to be at, or shortly before the oldest time in the array; however, depending on the action path, objects may not always be updated. If any object time stamp is updated outside of $\theta $ from another related object time stamp, then it can be inferred that a separate instance of the same action must have happened.

The function $SupportTest$ is defined that accepts an object update threshold and the $TraceStates$ array. First, the $TraceStates$ array is sorted based on the time stamp values, where element $0$ is the oldest and $n-1$ is the newest (most recent) time stamp value. Each object time stamp value is compared to the oldest time stamp value in the $TraceStates$ array. The comparison takes place until the time stamp value plus the object update threshold is less than the newest compared time stamp value in the array. When this happens, all time stamp values are assigned to the action instance that must have occurred between the oldest time stamp value and the most recent time stamp that is still less than the threshold. The oldest time stamp is then replaced with the most recent time stamp that is greater than the threshold, and the process starts again.

\begin{description}
\item \textbf{function} $SupportTest(\theta, TraceStates)$
\item      sort $TraceStates$
\item      $timeValue = TraceStates[0]$
\item      $foreach\ i\ in\ TraceStates;\ do$
\item         $if (timeValue + \theta > TraceStates[i-1, i-1])$
\item           $next$
\item      else
\item         $array detected = detected[] + [timeValue, \\TraceStates[i-2]]$
\item         let $timeValue = TraceStates[i-1]$
\item      done
\item return $detected[]$
\end{description}

\subsubsection{Shared Object Update Consistency}
Shared object time stamps are defined as \textit{a subset of time stamps $S\textsubscript{shared}$ in $T$ that may or may not be updated to the current value of the system clock on the occurrence of each execution of multiple actions}.

Shared object time stamps may be either `always updated' or `irregularly updated' category types depending on the action. Since the particular trace may be associated with more than one action, it is possible that any associated action instance could have updated the trace. Thus, without additional information, the only information that can be inferred from the detection of a shared object time stamps is that at least one of the associated actions must have happened.

An example of a shared object time stamp would be the access time stamp of a .dll file. Multiple actions can cause the .dll file to be accessed, so when examining the system in a post-mortem environment with no additional information, each action that causes the .dll file to be accessed has the same probability to have updated the accessed time stamp.

From this definition, an object time stamp update consistency function ($SharedTest$) can be derived to test whether each trace conforms to the shared object update category, and determine what action the trace is associated with. In the case of shared, if each trace has been updated within $\theta$, then the execution time for the action can be approximated to be at, or shortly before the oldest time in the array; however, objects may not always be updated. If any object is updated outside of $\theta$ from another object, then a separate execution of the same action may be inferred. Traces may also be associated with multiple actions. Because of this, additional context is needed to determine which action caused the trace. Some methods, such as probabilistic association, can be used to attempt to determine which action the trace is associated with, but there are limitations to these methods. However, action-to-trace association methods, and their weaknesses, are beyond the scope of this work.

Keeping with the currently defined signature creation model, at best what can be said when observing a shared trace is that all actions associated with the trace could have happened within their respective update thresholds. For this reason, the consistency checking of a group of shared objects is much like supporting objects. Each object time stamp value is compared from oldest to most recent. Each time stamp is considered to be associated with one action instance if the value is within the update threshold. The result is an array with multiple instances of the action. In the case of shared, the objects may be tested more than once, since they will be present in multiple signatures. This means that a trace may be associated with multiple actions, and all actions associated with the shared trace are assumed to have happened.

\begin{description}
\item \textbf{function} $SharedTest(\theta, TraceStates)$
\item sort $TraceStates$
\item $timeValue = TraceStates[0]$
\item $foreach\ i\ in\ TraceStates;\ do$
\item $if (timeValue + \theta > TraceStates[i-1, i-1])$
\item $next$
\item else
\item $array detected = detected[] + [timeValue, \\TraceStates[i-2]]$
\item $let timeValue = TraceStates[i-1]$
\item done
\item return $detected[]$
\end{description}

\subsection{Object Update Threshold}
The object update process is not instantaneous. In order to accurately differentiate between multiple instances of an action, object update duration must be defined for the particular action. The object update threshold, in seconds, of the actions ``Open Internet Explorer 8'' (IE8) and ``Open Firefox 3.6'' (FF3) were surveyed on 25 computer systems running Windows XP or Windows 7, and modeled as a normal distribution. To attempt to reduce noise, a limiter of 2$\sigma$ will be used.

\begin{figure}
	\caption{Normal distribution of IE8 execution times in seconds to two standard deviations.}
	\centering
	\includegraphics[width=0.9\linewidth]{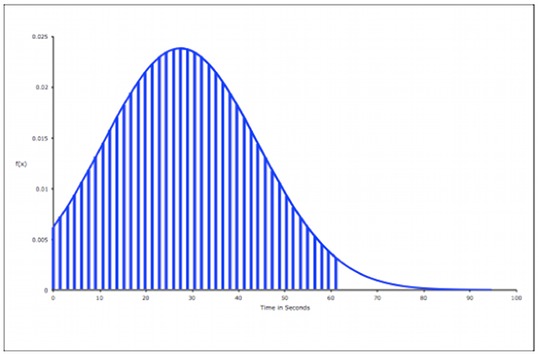}
	\label{fig3}
\end{figure}

\begin{figure}
	\caption{Histogram of IE8 execution times where X is time in seconds and Y is the number of occurrences within the update duration.}
	\centering
	\includegraphics[width=0.9\linewidth]{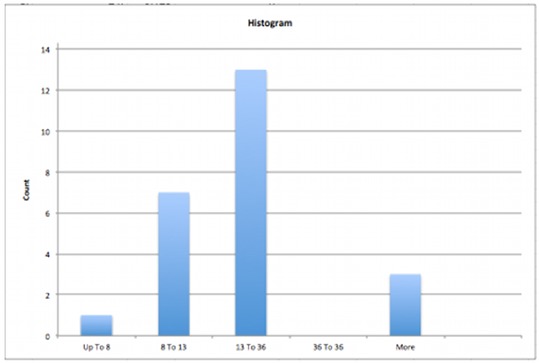}
	\label{fig4}
\end{figure}

For the action ``Open Internet Explorer 8'', the average execution update duration was 27.4 seconds, with a standard deviation ($\sigma$) of 16.76 seconds. The threshold chosen was 2$\sigma$, or from 0 to 61 seconds. Fig. \ref{fig3} shows the results modeled as a normal distribution, with a histogram of the data shown in Fig. \ref{fig4}.

\begin{figure}
	\caption{Normal distribution of FF3 execution times in seconds to two standard deviations.}
	\centering
	\includegraphics[width=0.9\linewidth]{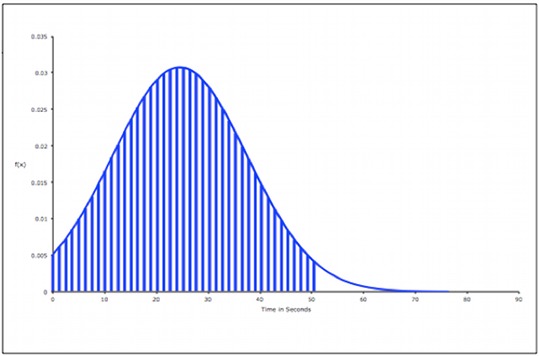}
	\label{fig5}
\end{figure}

\begin{figure}
	\caption{Histogram of FF3 execution times where X is time in seconds and Y is the number of occurrences within the update duration.}
	\centering
	\includegraphics[width=0.9\linewidth]{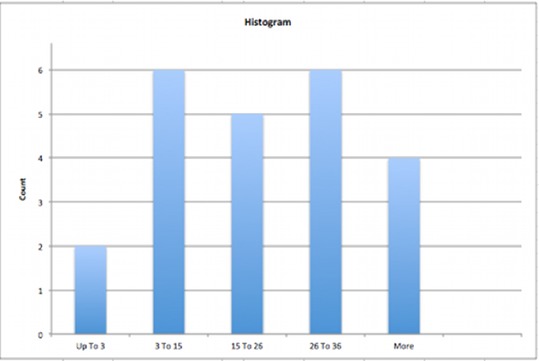}
	\label{fig6}
\end{figure}

For the action ``Open Firefox 3.6'', the average execution object update duration was 24.5 seconds, with a standard deviation of 12.96 seconds. The threshold chosen was 2$\sigma$, or from 0 to 50 seconds (Fig. \ref{fig5}, \ref{fig6}).

Objects will be associated to the same action instance if each has been updated within the given threshold. To handle the situation of an overlap of two action instances where an object could be in a position to be associated with either instance of the action, a search for the first and last time stamp in the set will be conducted, starting from the known point. If the update duration between the oldest object and the newest object is greater than the defined action's object update threshold, then at least two instances of the action must have happened. For example, an artifact $t_2$ is observed. $t_2$ is associated to an action whose signature consists of the set $\{t_1, t_2, t_3\}$, and whose threshold ($\theta $) is 60 seconds. If $t_1$ = 12:59:30, $t_2$ = 13:00:00, and $t_3$ = 13:00:58 then $t_2 - t_1 < \theta$ and $t_3 - t_2 < \theta$. In this case $t_1$ and $t_3$ could be associated to the same action instance since $\theta < 60$ when compared to $t_2$, even though $\theta < t_3 - t_1 = 88$. To handle this situation, when an object is found, all existing time stamps associated to the action instance are observed and sorted. The oldest time stamp is then used as the base from where all other returned time stamps are compared. If any time stamp is greater than $\theta$ from the oldest time stamp, then multiple instances of the action must have happened.

\subsection{Signature Analysis Model}
The proposed signature analysis model for detecting actions uses the previously defined classes of signatures in a layered approach to build up knowledge of actions that have happened in a system. To illustrate, a fictional example of this approach is given:

An action, $ActionX $, has a Core signature ($S\textsubscript{XCore}$) consisting of two time stamps, and a Supporting signature ($S\textsubscript{XSupport}$) that has three associated time stamps. All object update thresholds are defined as 30 seconds.

\begin{description} 
\item $S\textsubscript{XCore} = \{ [(o_1,t_1), (o_2,t_2)], 30sec., Core\}$
\item $S\textsubscript{XSupport} = \{[(o_3,t_3), (o_4,t_4), (o_5,t_5)], 30sec.,\\Supporting\}$
\end{description}

The Shared signature ($S\textsubscript{XShared}$) for $ActionX $ has two associated time stamps. Both of these time stamps are also associated with another action, $ActionY $. The Shared signature for $ActionY $ is denoted as $S\textsubscript{YShared}$.

\begin{description} 
\item $S\textsubscript{XShared} = \{[(o_6,t_6), (o_7,t_7)], 30sec., Shared\}$
\item $S\textsubscript{YShared} = \{[(o_6,t_6), (o_7,t_7)], 30sec., Shared\}$
\end{description}

The function $SignatureMatch(O',S)$ takes the system and signature as input, and returns the value of the observed action instance update time-span.

\begin{description} 
\item $\textbf{SignatureMatch}(O', S\textsubscript{XCore})$ returns
\item $\{[``4/14/2010\ 19:28:25'', ``4/14/2010\ 19:28:32'']\}$
\item $\textbf{SignatureMatch}(O', S\textsubscript{XSupport})$ returns
\item $\{[``4/14/2010\ 15:13:25''], [``4/14/2010\ 19:28:18'', ``4/14/2010\ 19:28:34'']\}$
\item $\textbf{SignatureMatch}(O', S\textsubscript{XShared})$ returns
\item $\{[``4/14/2010\ 19:28:25''], [``5/2/2010\ 9:45:02'']\}$
\item $\textbf{SignatureMatch}(O', S\textsubscript{YShared})$ returns
\item $\{[``4/14/2010\ 19:28:25''], [``5/2/2010\ 9:45:02'']\}$
\end{description}
The result of this detection process is summarized in Table \ref{tab1}.

\begin{table}
\centering
\begin{tabular}{|l|l|l|}
\hline
~ & ActionX & ActionY \\\hline
Core & 4/14/2010 19.28:25 & ~\\\hline
Core & 4/14/2010 19:28:32 & ~\\\hline
Support & 4/14/2010 15:13:25 & ~\\\hline
Support & 4/14/2010 19:28:18 & ~\\\hline
Support & 4/14/2010 19:28:34 & ~\\\hline
Shared & 4/14/2010 19:28:25 & 4/14/2010 19:28:25\\\hline
Shared & 05/02/2010 09:45 & 05/02/2010 09:45\\\hline
\end{tabular}
\caption{Summary of example time stamps related to $ActionX$.}
\label{tab1}
\end{table}

Since Core signature traces are always updated and relate only to $ActionX$, it can be inferred that $ActionX$ last happened approximately at 4/14/2010 19:28:25. Both $ActionX$ Core timestamps are within $\theta$, so the traces are consistent.

With the knowledge of the last execution time of $ActionX$, the Supporting signature may now provide more information. In this case, two supporting traces confirm the last execution time ($t_3$ and $t_4$). Traces in the Supporting signature may not always be updated, as is shown by the supporting trace ($t_5$) with a time stamp of 4/14/2010 15:13:25. This trace is consistent since the time is before the identified last execution time ($t_1$). Also, since Supporting traces are associated only with one action, a previous execution of $ActionX$ must have happened at this time.

Finally, Shared traces are examined. Each trace is associated with both $ActionX$ and $ActionY$. The first shared trace ($t_6$) has a time stamp that is within the last execution time of $ActionX$; however, $ActionY$ could have also happened at this time. Calculating the probability of one trace belonging to a particular action has been discussed in \cite{Carney2004,Kwan2008}, but is beyond the scope of this work. Because of this, no conclusion can be made. The next trace, however, has a time that is after the detected last execution time ($t_1$) of $ActionX$. Since this trace is associated only with $ActionX$ or $ActionY$, it can be inferred that the trace ($t_7$) must belong to $ActionY$ since it is not consistent with the information known about $ActionX$. An instance of $ActionY$ must have happened at approximately 5/2/2010 9:45:02, to be consistent with $ActionX$.

After this analysis, action instance approximations may be given as shown in Table \ref{tab2}.

\begin{table}
\centering
\begin{tabular}{|l|p{2cm}|p{2cm}|}
\hline
~ & ActionX & ActionY \\\hline
Last Execution & 4/14/2010 19.28:18 & ~ \\\hline
Previous Execution & 4/14/2010 15:13:25 & 05/02/2010 09:45 \\\hline
\end{tabular}
\caption{Known action instance approximation times after signature analysis.}
\label{tab2}
\end{table}

\begin{figure}
	\caption{Graph of objects in $T$ related to $ActionX$ grouped by $\theta $, that shows two distinct instances of $ActionX$.}
	\centering
	\includegraphics[width=0.45\textwidth]{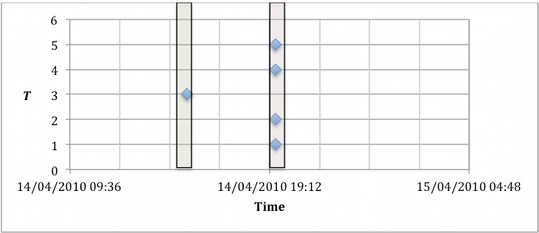}
	\label{fig7}
\end{figure}

The time stamps that are known to relate only to $ActionX$ are shown in Fig. \ref{fig7}. The times are grouped, where $\theta $ = 30 seconds. In the case of Core and Supporting signatures, where the traces are related only to $ActionX$, the most recent, as well as past executions of the action can be inferred.

This example illustrates that by layering multiple observations more information about previous executions of actions can be automatically inferred. Also, by building on already detected information, inferences about other non-related actions may be made. Evaluation of this method will be presented in a case study, where the process is applied to detect actions in a real environment.

\section{Case Study}
To test the proposed method, example signatures were created for the actions ``Open Internet Explorer 8'' and ``Open Firefox 3.6'' on Windows XP. Windows XP was chosen since, at the time of data collection, it was still the most frequently encountered operating system of surveyed law enforcement \cite{James2010b}. For brevity, only Core and Supporting objects will be analyzed. The tested signatures are defined as regular expressions\footnote{For more information on Regular Expressions, see http://www.bsd.org/regexintro.html} to support portability, and are listed in Table \ref{tab3} and Table \ref{tab4}. As previously shown, $\theta$ for FF3 is 50 seconds, and $\theta$ for IE8 is 61 seconds.

\begin{table}
\centering
\begin{tabular}{|p{1.25cm}|p{1.6cm}|p{4.3cm}|}
\hline
Category & Time Stamp & Objects related to `Opening FF3'\\\hline
Core & Modified & .*/Firefox/Profiles/.*default/
urlclassifierkey.{\textbackslash}.txt\\\hline
Core & Modified & .*/Prefetch/Firefox{\textbackslash}.EXE-.*{\textbackslash}.pf\\\hline
Support & Created & .*/Prefetch/Firefox{\textbackslash}.EXE-.*{\textbackslash}.pf\\\hline
Support & Created & .*/Firefox/Profiles/
.*default/cookies.sqlite-journal\\\hline
Support & Created & .*/Firefox/Profiles/
.*{\textbackslash}default/urlclassifierkey.{\textbackslash}.txt\\\hline
Support & Created & .*/Firefox/Profiles/
.*default/startupCache\$\\\hline
Support & Created & .*/Firefox/Profiles/
.*default/pluginreg.dat\\\hline
\end{tabular}
\caption{FF3 objects, categories and corresponding time stamp of interest.}
\label{tab3}
\end{table}

\begin{table}
\centering
\begin{tabular}{|p{1.25cm}|p{1.6cm}|p{4.3cm}|}
\hline
Category & Time Stamp & Objects related to `Opening IE8'\\\hline
Core & Modified & .*/Prefetch/
IEXPLORE{\textbackslash}.EXE-.*{\textbackslash}.pf\\\hline
Support & Created & .*/Prefetch/
IEXPLORE{\textbackslash}.EXE-.*{\textbackslash}.pf\\\hline
Support & Created & .*/Cookies/.*@ATDMT{\textbackslash}[[0-9]{\textbackslash}]{\textbackslash}.TXT\\\hline
Support & Created & .*/Cookies/.*@BING{\textbackslash}[[0-9]{\textbackslash}]{\textbackslash}.TXT\\\hline
Support & Created & .*/Cookies/.*@live{\textbackslash}[[0-9]{\textbackslash}]{\textbackslash}.TXT\\\hline
\end{tabular}
\caption{IE8 objects, categories and corresponding time stamp of interest.}
\label{tab4}
\end{table}

The case consists of two computers running Windows XP with both IE8 and FF3 installed. Each computer was used daily for entertainment, work and study tasks. Both users identified that they used Firefox as their primary browser. To accurately determine when IE8 and FF3 have been opened and closed, a Windows Security auditing policy was implemented on both computers to monitor process creation and executable access. Each computer was monitored for a number of days, after which the Windows security event log was exported and the computer's file system meta-data was collected with tools from The Sleuth Kit version 3.2.2. The resulting Windows Security Logs and meta-data outputs are available as a downloadable dataset \cite{James2013}.

On `Computer 1', 12 instances of opening FF3 from the 19th to the 24th were identified from the Windows event log, and 6 instances of opening IE8 were identified. The meta-data from Computer 1 was scanned using the previously defined signature for opening FF3. The identified objects and associated time stamps are shown in Table \ref{tab5}.

In this case, all Core objects were discovered. However, one Core object had a time stamp that was different than another Core object. This unexpected behavior can be explained by looking at the Open and Close times from the Windows event log. The Firefox open event with the process ID 4284 occurred at 13:23, and was never followed by a process close event. While process 4284 was still open, another instance of Firefox was started, process 5480, at 15:02. If process 4284 had locked the object in question, then the time stamp may not be updated upon another instance of the action. However, $T_2$ was not locked by the first event, and was updated. Since both objects must be updated when the action happens, this must mean that two instances of the same action must be running in parallel, otherwise both artifacts would be updated.

\begin{table}
\centering
\begin{tabular}{|p{1.25cm}|p{1.6cm}|p{4.3cm}|}
\hline
Category & Time & Returned Object\\\hline
Core & 07/24/2011 13:24:14 & C:/Documents and Settings/User1/
Application Data/Mozilla/Firefox/
Profiles/94370b5u.default/
urlclassifierkey3.txt\\\hline
Core & 07/24/2011 15:02:31 & C:/Windows/Prefetch/
Firefox.exe-28641590.pf\\\hline
Support & 12:26/2010 04:26:24 & C:/Windows/Prefetch/
Firefox.exe-28641590.pf\\\hline
Support & 07/24/2011 13:24:10 & C:/Documents and Settings/User1/
Application Data/Mozilla/Firefox/
Profiles/94370b5u.default/
cookies.sqlite-journal\\\hline
Support & 01/05/2011 23:15:34 & C:/Documents and Settings/User1/
Application Data/Mozilla/Firefox/
Profiles/94370b5u.default/
urlclassifierkey3.txt\\\hline
Support & N/A & .*/Firefox/Profiles/
.*default/startupCache\$\\\hline
Support & 12/26/2010 03:04:55 & C:/Documents and Settings/User1/
Application Data/Mozilla/Firefox/
Profiles/94370b5u.default/
pluginreg.dat\\\hline
\end{tabular}
\caption{FF3 objects and associated timestamps found using signature detection on Computer 1.}
\label{tab5}
\end{table}

Next, the meta-data from Computer 1 was scanned using the previously defined signature for opening IE8. The identified objects and associated time stamps are shown in Table \ref{tab6}. In this case, all Core objects were detected, identifying the most recent execution of IE8 as happening at approximately 14:56 on 07/23/2011. All other associated objects had timestamps before this time. 

\begin{table}
\centering
\begin{tabular}{|p{1.25cm}|p{1.6cm}|p{4.3cm}|}
\hline
Category & Time & Returned Object\\\hline
Core & 07/23/2011 14:56:53 & C:/Windows/Prefetch/
Iexplore.exe-27122324.pf\\\hline
Support & 07/19/2011 00:57:22 & C:/Windows/Prefetch/
Iexplore.exe-27122324.pf\\\hline
Support & 06/14/2011 10:47:26 & C:/Documents and Settings/User1/
Cookies/user1@atdmt[2].txt\\\hline
Support & 01/11/2011 19:40:26 & C:/Documents and Settings/User1/
Cookies/user1@bing[2].txt\\\hline
Support & 06/14/2011 10:47:26 & C:/Documents and Settings/User1/
Cookies/user1@live[1].txt\\\hline
\end{tabular}
\caption{IE8 objects and associated time stamps found using signature detection on Computer 1.}
\label{tab6}
\end{table}

On Computer 2, 14 instances of opening FF3 from the 13th to the 17th was identified from the Windows event log, and two instances of opening IE8 were identified. The meta-data from Computer 2 was scanned using the previously defined signature for opening FF3. The identified objects and associated time stamps are shown in Table \ref{tab7}. In this case all Core artifacts were detected, identifying the most recent execution of FF3 as approximately 20:24 on 07/17/2011. All other associated objects had time stamps before this time. 

\begin{table}
\centering
\begin{tabular}{|p{1.25cm}|p{1.6cm}|p{4.3cm}|}
\hline
Category & Time & Returned Object\\\hline
Core & 07/17/2011 20:24:26 & C:/Documents and Settings/user/
Application Data/Mozilla/Firefox/
Profiles/c2yzki95.default/
urlclassifierkey3.txt\\\hline
Core & 07/17/2011 20:24:18 & C:/Windows/Prefetch/
Firefox.exe-28641590.pf\\\hline
Support & 05/21/2010 16:15:23 & C:/Windows/Prefetch/
Firefox.exe-28641590.pf\\\hline
Support & 05/14/2010 16:44:22 & C:/Documents and Settings/user/
Application Data/Mozilla/Firefox/
Profiles/c2yzki95.default/
cookies.sqlite-journal\\\hline
Support & 10/23/2010 11:26:02 & C:/Documents and Settings/user/
Application Data/Mozilla/Firefox/
Profiles/c2yzki95.default/
cookies.sqlite-journal (deleted)\\\hline
Support & 04/13/2011 00:33:49 & C:/Documents and Settings/user/
Application Data/Mozilla/Firefox/
Profiles/c2yzki95.default/
urlclassifierkey3.txt\\\hline
Support & 07/17/2011 15:23:05 & C:/Documents and Settings/user/
Application Data/Mozilla/Firefox/
Profiles/c2yzki95.default/
startupCache\\\hline
Support & 07/17/2011 00:46:36 & C:/Documents and Settings/user/
Application Data/Mozilla/Firefox/
Profiles/c2yzki95.default/
pluginreg.dat\\\hline
\end{tabular}
\caption{FF3 objects and associated time stamps found using signature detection on Computer 2.}
\label{tab7}
\end{table}

Next, the meta-data from Computer 2 was scanned using the previously defined signature for opening IE8. The identified objects and associated time stamps are shown in Table \ref{tab8}. In this case, all Core artifacts were detected, setting the most recent execution of IE8 at approximately 15:15 on 07/17/2011. All other associated objects had time stamps before this time.

\begin{table}
\centering
\begin{tabular}{|p{1.25cm}|p{1.6cm}|p{4.3cm}|}
\hline
Category & Time & Returned Object\\\hline
Core & 07/17/2011 15:15:13 & C:/Windows/Prefetch/
Iexplore.exe-27122324.pf\\\hline
Support & 07/17/2011 15:15:09 & C:/Windows/Prefetch/
Iexplore.exe-27122324.pf\\\hline
Support & 03/10/2011 15:01:01 & C:/Documents and Settings/User1/
Cookies/user1@atdmt[1].txt\\\hline
Support & 03/10/2011 15:38:37 & C:/Documents and Settings/User1/
Cookies/user1@bing[2].txt\\\hline
Support & 03/10/2011 15:38:37 & C:/Documents and Settings/User1/
Cookies/user1@live[2].txt\\\hline
\end{tabular}
\caption{IE8 objects and associated time stamps found using signature detection on Computer 2.}
\label{tab8}
\end{table}

\subsection{Evaluation}
In this case study signatures were used to detect Core and Supporting objects. After, objects within the action's object update threshold were grouped and considered related to the same action instance. The result is a list of object timestamps related to a particular action of a particular computer. The results of the previous signature detection, where the Windows Event log confirms all detected timestamps, are given in Table \ref{tab9}.

Table \ref{tab9} shows that the most recent execution, as well as at least one past instance of Firefox was detected on both computers. Further, at least the most recent execution of Internet Explorer was detected on both computers.

Detecting more previous instances of Firefox than Internet Explorer may be a result of both users using Firefox as their primary browsers, meaning that there may have been fewer instances of Internet Explorer during the experiment. In all cases, however, the most recent instance of the action was always accurately detected.

\begin{table}
\centering
\begin{tabular}{|p{1.8cm}|p{1cm}|p{1.8cm}|p{1.8cm}|}
\hline
Computer & Action & Logged Time & Detected Time\\\hline
Computer 1 & Open FF3 & 07/24/2011 15:02:30 & 07/24/2011 15:02:31\\\hline
Computer 1 & Open FF3 & 07/24/2011 13:23:58 & 07/24/2011 13:24:14\\\hline
Computer 2 & Open FF3 & 07/17/2011 20:24:14 & 07/17/2011 20:24:18\\\hline
Computer 2 & Open FF3 & 07/17/2011 15:23:01 & 07/17/2011 15:23:05\\\hline
Computer 2 & Open FF3 & 07/17/2011 00:46:14 & 07/17/2011 00:46:36\\\hline
Computer 1 & Open IE8 & 07/23/2011 14:56:46 & 07/23/2011 14:56:53\\\hline
Computer 2 & Open IE8 & 07/17/2011 15:15:06 & 07/17/2011 15:15:09\\\hline
\end{tabular}
\caption{Detected executions of an action based on signature matching over meta-data separated by a measured object update threshold, where the Windows Security Log confirms each detected time.}
\label{tab9}
\end{table}

\subsubsection{Further Implementation}
This case study was meant to illustrate the described theory. The researchers are currently implementing the proposed model in a tool designed for digital investigators, and are applying it to anti-forensic action detection\footnote{The open source tool implementing the proposed theory can be found at http://github.com/hvva/IoAF}. For signature creation, specific actions of interest are chosen, and a sandbox is used to extract object updates in a target system. Object update `traces' are categorized based on the given	 model, and relevant traces are added as signatures for actions of interest. The tool then accepts a suspect disk image (or live computer), and matches generated signatures with extracted suspect meta-data to reconstruct action instances in the suspect system.

\subsubsection{Weaknesses}
The greatest weakness with this method is the same weakness in all signature-based detection methods. Not all possible actions can be known, and unknown actions will not be considered. This lack of complete knowledge is especially relevant in the case of action instance signature generation. If not all possible actions are known, then it is impossible definitely determine a trace's category. The answer to this weakness comes from the investigator. Human knowledge is also incomplete, yet human investigators are able to state that actions in a system must have happened based on their knowledge of how the system works. Human investigators also update their knowledge based on new experiences. Signature-based action instance detection methods must be able to be updated when new information that leads to changing a trace's category is found. Even if action signatures are being updated, it is still not possible to account for every piece of custom-made software. For this reason, this method is more suitable as a pre-analysis inference guide for the human investigator or for post-examination human inference verification rather than for completely automated investigations.

Another consideration is when an incident has multiple relevant actions. While some actions can be differentiated by using Core, Supporting and Shared objects, if relevant actions happen within a very short time of each other -- as the formal model shows -- relating the object to the specific action is impossible. The final state of the system simply does not contain enough information to differentiate two similar action instances happening at approximately the same time. Further, this method can help detect action instances, but in most cases it will not help with the problem of attribution, unless a specific, attributable pattern is detectable. For example if malicious software creates detectable object update patterns in the system.

\section{Conclusions}
In this work, the concept of signature detection of actions was discussed. Three categories of signatures were formally defined that, based on their unique update patterns, allow for the detection of the most recent as well as past action instance approximations. A method for differentiating between two instances of the same action within a very short time range was also given. After, the signature analysis method was illustrated with an example. Finally, a case study using the proposed signature analysis method on two real-world computers was examined. The case study showed good results in detecting the most recent instance of the action, and did give more information about past action instances. However, the method did not detect every past instance of the action, but it also did not give false positives. Overall, by using the proposed method to automatically approximate action instances and display them in a timeline, an investigator can very quickly get an idea of how, and when, the system has been used regardless of their knowledge of the system or specific data being analyzed.

Future work will look at improving the action instance extraction by integrating more objects/content into the signature, such as Windows Registry and system log entries. Further, much like intrusion detection systems, a hybrid statistical and signature-based approach may help reduce signature-based weaknesses while still maintaining a high level of accuracy. And finally, future work will attempt to monitor a system for a longer period of time to determine if detected times very far in the past correspond with actual action instances.

\bibliographystyle{unsrt}
\bibliography{bibfile}

\end{document}